# Icosahedral quasicrystal and 1/1 cubic approximant in Au-Al-Yb alloys

Tsutomu Ishimasa, Yukinori Tanaka and Shiro Kashimoto

*Graduate School of Engineering, Hokkaido University, 060-8628 Sapporo, Japan*

Abstract

A P-type icosahedral quasicrystal is formed in Au-Al-Yb alloy of which 6-dimensional lattice parameter $a_{6D}$ = 7.448 Å. The composition of the quasicrystal was analyzed to be $Au_{51}Al_{34}Yb_{15}$. This quasicrystal is formed in as-cast alloys, and is regarded as metastable because of decomposition into other crystalline phases by annealing at 700 ºC. Among Tsai-type quasicrystals, this quasicrystal is situated just between Zn-Sc group with smaller $a_{6D}$ and larger Cd-Yb group. Intermediate valence of Yb recently observed in this quasicrystal may be due to this unique situation, namely smaller major component Au-Al than in Cd-Yb.

The predominant phase in the annealed specimen is a 1/1 cubic approximant with lattice parameter $a$ = 14.500 Å belonging to the space group $Im\bar{3}$. This phase is stable at the composition $Au_{51}Al_{35}Yb_{14}$ at 700 ºC. Rietveld structural analysis indicated that the crystal structure is understood as periodic arrangement of Tsai-type clusters each including four Au-Al atoms at their centers. Chemical ordering of Au and Al is characteristics of this approximant.

*Corresponding author
  E-mail: ishimasa@eng.hokudai.ac.jp
  Telephone: +81-11-706-6643
  Fax: +81-11-706-6859

1. **Introduction**

   After the discovery of binary stable Cd-Yb quasicrystal [1, 2], the icosahedral quasicrystals containing Yb have been found in several alloy systems such as Zn-Mg-Yb [3], Cd-Mg-Yb [4], Ag-In-Yb [5] and Au-In-Yb [6]. These quasicrystals as well as the corresponding approximant crystals have been studied from the view points of metallurgy, crystallography and solid-state physics. In particular the valence of Yb ions in these compounds is of interest in relation to possible occurrence of intermediate-valence state. However, in all the cases there had been no indication of intermediate-valence state yet. Then the recent report by Kawana *et al*. [7] seems promising. They have reported appearance of intermediate-valence state of Yb both in Cd-Yb quasicrystal and its 1/1 approximant under high pressure. This may be related to the fact that the atomic size of trivalent Yb is smaller than that of divalent Yb. Similar evolution of trivalent state has been observed at ambient pressure by substitution, for example in the case of Laves phase $Yb_xIn_{1-x}Cu_2$ [8] as well as in $YbCu_{5-x}Ga_x$ [9]. In the present study, new quasicrystal alloy containing Yb has been searched in order to realize such situation. Here the substitution rule empirically known in Tsai-type quasicrystals was applied. Common properties of Tsai-type quasicrystals are summarized as follows:

   Remarkable feature of Tsai-type quasicrystals is variety in base metals, namely Zn [10], Zn-M (M = Mg, Cu, Pd, Ag, Pt, Au) [3, 11-13], Zn-T (T: Mn, Fe, Co, Ni) [14], Cu-Al [15], Cu-Ga [16, 17], Cd [1, 2], Cd-Mg [4, 18], Ag-Al [19], Ag-In [5], Pd-Al [20], Au-Ga [19, 21] and Au-In [6, 22]. As minor component they contain trivalent transition elements such as Sc, Y or lanthanoids, or divalent Ca or Yb. In all the cases the composition of the minor component ranges from 12 to 16 at %, and then Tsai-type quasicrystals satisfy almost constant electron concentration *e*/*a* ranging from 2.00 to 2.15. This is the first condition of the famous Hume-Rothery rules. According to magnitude of 6-dimensional lattice parameters $a_{6D}$, these quasicrystals are classified into two groups; Zn-Sc group with smaller $a_{6D}$ and Cd-Yb group with larger one. For the first group the minor component is always Sc, and for the second group that is larger element such as Ca, Y or lanthanoids. Such combination may be due to geometrical condition for the construction of Tsai-type cluster with concentric triple shells, and induces the second condition of Hume-Rothery rules; size factor [3, 23]. The two conditions on *e*/*a* and size factor can be regarded as substitution rule in Tsai-type quasicrystals.

An approximant is a periodic structure consisting of similar local structural unit, namely cluster, to the corresponding quasicrystal. With respect to constituent elements and compositions, approximants obey very similar formation conditions to those of quasicrystals, while the difference that governs formation of either quasicrystal or approximant has not yet been clarified. Study of approximants has given us useful hint to understand structural and physical properties of quasicrystals. In the case of Tsai-type, close structural relationship between $Cd_6Yb$ ($Cd_6Ca$) approximant and the quasicrystals was discussed from the view points of three-dimensional Penrose tiling [24]. Now we know that the structure of a 1/1 approximant is understood as periodic projection of a six-dimensional crystal from which a structural model of the quasicrystal is constructed. This idea was effectively used in the structural analysis of the Cd-Yb quasicrystal [25].

In this paper, experimental results on replacement of Cd by Au-Al will be reported. It will be shown that from the structural view points new Au-Al-Yb quasicrystal has unique position between the two structure groups; namely Zn-Sc and Cd-Yb groups. The result of structural analysis of 1/1 approximant in Au-Al-Yb alloy will also be described in details.

## 2. Experimental

Alloy ingots of $Au_xAl_{100-x-y}Yb_y$ were synthesized in the narrow composition region of $44 \leq x \leq 50$ and $15 \leq y \leq 17$ referring to the stoichiometry of $Cu_{46}Al_{38}Sc_{16}$ quasicrystal [15]. As-cast specimens were prepared using an arc furnace in an atmosphere of Ar. They were further annealed at 700 ºC for 24 h in order to study stability of phases. Before the annealing, the specimen was sealed in an silica ampoule after evacuating to pressure approximately $2 \times 10^{-6}$ Torr.

The compositions of phases formed in the Au-Al-Yb alloys were analyzed by means of electron probe micro-analysis (EPMA) using a micro-analyzer JXA-8900M equipped with a wavelength-dispersive type spectrometer at an acceleration voltage 20 kV. The electron microscopic observation was carried out at 200 kV by a JEOL JEM200CS microscope equipped with a double tilting stage. An alloy ingot was crushed by agate mortar and pestle, and thin part was observed by the electron microscope.

The powder X-ray diffraction experiments were carried out using a diffractometer

RINT-2500 equipped with a Cu X-ray tube and a pyrolytic graphite counter monochrometer. The radius of the diffractometer is 185 mm. In the case of 1/1 approximant, the following attentions were paid in order to obtain reliable diffraction data for Rietveld analysis using the conventional diffractometer: A narrow divergence slit with 0.25º was used to measure correct intensity down to $2\theta = 6º$. The sample plate was rotated around the axis normal to the plate in order to ensure randomness of crystal orientation. Intensity at each point was measured for 30 s in order to reduce statistical noise. By this procedure intensity data with the interval of 0.02 º were collected in the angle range $6 º \leq 2\theta \leq 146 º$. The intensity distribution was analyzed by the Rietveld method using the software package RIETAN-2000 developed by Izumi [26].

## 3. Results
### 3.1. *Icosahedral quasicrystal and 1/1 approximant in Au-Al-Yb alloys*

An icosahedral quasicrystal was observed almost exclusively in the as-cast specimen with the nominal composition $Au_{49}Al_{34}Yb_{17}$. X-ray diffraction pattern of the alloy is presented in figure 1. All the peaks were indexed as a P-type icosahedral quasicrystal with the 6-dimensional lattice parameter $a_{6D}$ = 7.448 (2) Å except for a small peak indicated by an arrow in figure 1. Electron diffraction patterns presented in figure 2 also confirms the formation of the quasicrystal, although weak reflections tend to deviate from the exact symmetric positions as can be seen in figure 2a. The composition of the quasicrystal was analyzed to be $Au_{51}Al_{34}Yb_{15}$ by EPMA. While the quasicrystal was observed also in the as-cast $Au_{47}Al_{37}Yb_{16}$ and $Au_{48}Al_{36}Yb_{16}$ specimens, no quasicrystal has been detected in all the specimens annealed at 700 ºC. This result suggests metastability of the quasicrystal. Instead of the quasicrystal three other phases were detected in the annealed $Au_{49}Al_{34}Yb_{17}$ specimen by EPMA. Their analyzed compositions are $Au_{51}Al_{35}Yb_{14}$, $Au_{46}Al_{36}Yb_{18}$ and $Au_{56}Al_{27}Yb_{17}$ respectively. The first phase was assigned to a 1/1 cubic approximant.

The 1/1 approximant was observed as a predominant phase in the following six specimens annealed at 700 ºC with nominal compositions $Au_{49}Al_{36}Yb_{15}$, $Au_{47}Al_{37}Yb_{16}$, $Au_{48}Al_{36}Yb_{16}$, $Au_{49}Al_{35}Yb_{16}$, $Au_{50}Al_{34}Yb_{16}$, and $Au_{49}Al_{34}Yb_{17}$. In the first specimen the approximant formed exclusively as shown in figure 3. Electron diffraction experiments of this phase revealed that the Laue class is $m\bar{3}$ and the cubic lattice is body-centered type with the lattice parameter $a$ = 14.500 Å (See figure 4). These observations agree with the known properties of a 1/1 cubic approximant of Tsai-type [23]. The lattice parameter agrees well with the calculated one 14.497 Å using the 6-dimensional lattice parameter $a_{6D} = 7.448$ Å as 1/1. This agreement as well as the resemblance in the intensity distribution in diffraction patterns support that both the quasicrystal and the approximant include similar local structural unit, namely Tsai-type cluster, described below.

### 3.2. *Rietveld analysis of 1/1 approximant*

Rietveld analysis of the 1/1 approximant was carried out by assuming the space group $Im\bar{3}$ consistent to the above observation of the electron diffraction patterns. As the starting structure model, the Larson and Cromer model [27] proposed for $Cd_6Y$ was used. In the initial model all Cd atoms were replaced by an average atom $Au_{57.6}Al_{42.4}$, and Y by Yb. The ratio between Au and Al in each site was changed step by step in order to achieve better fit to the measured intensity distribution. In the final stage of the refinement, three possible structures were tested for the central region of Tsai-type cluster, namely partially occupied 16$f$, 24$g$ and 48$h$ sites. The first two correspond to $Cd_6Yb$ model [28] and Larson and Cromer model [27], respectively. Additional atoms at 8$c$ site, namely at [1/4, 1/4, 1/4] were also tested, which were indicated in the recent structure analyses of other 1/1 approximants [21, 22, 29]. The best fit with $R_{wp}$ = 5.86 %, $R_I$ =1.03 % and $R_F$ = 0.68% was obtained using both 48$h$ site and 8$c$ site. The result is summarized in Table 1. The lattice parameter was determined to be $a$ = 14.500 (2) Å by this analysis.

### 3.3. *Description of structure of 1/1 cubic approximant*

The present model consists of nine crystallographic sites listed in table 1. Characteristic feature of the model is chemical ordering between Au and Al beside the ordering of Yb that occupies exclusively the 24$g$ site forming an icosahedron in figure 5c. Other sites are occupied by mixed atoms of Au and Al except for the site named M8. The site M8 is occupied by Al with the occupancy 92% without Au. The mixed sites are classified into three groups; site containing Au more than 72 %, site containing Al more than 97 % and intermediate between them. Hereafter they will be referred as Au-site, Al-site and mixed site, respectively. The unit cell includes 89.6 Au, 61.8 Al and 24.0 Yb, and thus 175.4 atoms in total. The composition of the model is $Au_{51.1}Al_{35.2}Yb_{13.7}$ that agrees with the composition $Au_{51}Al_{35}Yb_{14}$ analyzed by EPMA as well as the nominal one $Au_{49}Al_{36}Yb_{15}$. The small reduction of Yb in the analyzed compositions from the nominal one suggests evaporation of Yb during the sample preparation.

The structure of the 1/1 approximant can be regarded as periodic arrangement of Tsai-type clusters with chemical ordering. Figure 5e presents a part of triacontahedron on which atoms M5 (Al-site) and M3 (Au-site) are arranged alternatively. The arrangement of this polyhedron forms a cage network with holes both at the origin and the body-center. Tsai-type clusters are embedded in these holes. The first shell of Tsai-type cluster is a dodecahedron composed of M2 (mixed site) and M4 (Au-site) shown in figure 5b. The second shell is an icosahedron of Yb. The third is an icosidodecahedron composed of M1 (Au-site) and M6 (mixed site) presented in figure 5d. At the center of the cluster four atoms in total are included, which belong to M7 (mixed site) with the occupancy 16.7 %. This complicated shape reflects an average of variously-oriented tetrahedrons as discussed by Gómez and Lidin [29]. The ordering scheme, namely preferential occupation by atoms, is very similar to that in $Au_{50.5}Ga_{35.9}Ca_{13.6}$ approximant reported by Lin and Corbett [21] as shown in the last column in Table 1.

4. Discussion

Figure 6 illustrates the other type of description of the structure as the framework of Yb atoms. This framework is regarded as a 3-dimensional tiling of four kinds of bricks; an icosahedron, an octahedron, a rhombus prism and a pyramid shown in figure 6a. Each brick has unique decoration with atoms presented in figures 6b ~ e. In the cubic unit cell there are two icosahedrons, eight octahedrons, six rhombus prisms and twelve pyramids. The icosahedron in figure 6d just corresponds to that in figure 5c. Each icosahedron is surrounded by eight octahedrons, six rhombus prisms, and twelve pyramids so as to form the Tsai-type cluster. The icosahedrons are connected in the 3-fold direction through the octahedron, and in the 2-fold direction through the rhombus prism. In the octahedron the arrangement of atoms looks like that in a body-centered cubic structure. The pyramid in figure 6c resembles a half of the octahedron.

It may be intriguing to compare the bricks with those in other approximant. In 1/0-2/1-1/0 approximant formed in Cu-Al-Sc alloy [30], there are similar icosahedron and octahedron. Besides them this approximant includes other three kinds of bricks which are not included in the present 1/1 approximant. In both approximants, the position vectors of twelve rare-earth atoms forming the icosahedron can be assigned by suitable projection to twelve fundamental vectors $\pm a_{6D}\boldsymbol{e}_i$ of six-dimensional hypercubic lattice, where $i = 1 \sim 6$ (for details see figure 3 in [30]). Furthermore in both approximants all the positions of the rare-earth atoms can be expressed by the projection of lattice vectors, namely linear combination of the fundamental vectors with integral coefficient. This fact suggests possible way to understand the structure of the Tsai-type quasicrystal; quasiperiodic framework of rare-earth atoms with icosahedrons plus decoration of each brick with chemical ordering similar to that in the corresponding approximant. This approach resembles that used for the construction of the structure model of Cd-Yb quasicrystal [25], but does not necessarily insist the frequent appearance of larger Tsai-type cluster with outer triacontahedron. It is necessary to check the validity of such approach through calculation of diffraction intensity.

Characteristic feature of the Au-Al-Yb quasicrystal can be noticed in figure 7a. In this figure the relation between $a_{6D}$ and atomic radius $\bar{r}$ is plotted for known Tsai-type quasicrystals. Here atomic radius $\bar{r}$ was calculated by averaging radii of constituent elements using the values listed in Table 4-4 in reference [33]. The linear relationship between $a_{6D}$ and $\bar{r}$ in figure 7a indicates that these quasicrystals are essentially isostructures belonging to a unique structure type, Tsai-type. In this diagram Tsai-type quasicrystals are classified into four groups: group 1 consisting of Cu-Sc alloys, group 2 of Zn-Sc alloys, group 3 including mainly Cd-Mg-L (L: lanthanoid), and finally group 4 of Cd-Yb and Cd-Ca alloys. Roughly speaking they can be classified into two groups: quasicrystals containing smaller Sc as minor component and those with larger Ca or lanthanoids. There is a wide gap between these groups ranging from $a_{6D}$ = 7.2 to 7.8 Å.

There are three exceptions which fall into this gap. They are $Zn_{76}Mg_{10}Yb_{14}$ [3], $Au_{50.3}Ga_{34.5}Ca_{15.2}$ [21] and $Au_{51}Al_{34}Yb_{15}$ discovered in this work. These special alloys are combinations of smaller major component and larger minor component. In the case of $Zn_{76}Mg_{10}Yb_{14}$, Yb is known to be divalent from magnetic measurement, and then for the calculation of $\bar{r}$ the radius of divalent Yb (1.940 Å) was used. While the $Zn_{76}Mg_{10}Yb_{14}$ just obeys the linear relationship, $Au_{50.3}Ga_{34.5}Ca_{15.2}$ and $Au_{51}Al_{34}Yb_{15}$ are slightly deviated. In the case of $Au_{51}Al_{34}Yb_{15}$ if one uses the radius of trivalent Yb (1.740 Å) for the calculation of $\bar{r}$, the plot is located on the left hand side of the line. Contrary, the radius of divalent Yb is too large. The intermediate radius seems appropriate to satisfy the linear relationship.

Similar linear relationship is also found for 1/1 approximants as can be seen in figure 7b. Corresponding to the diagram of the quasicrystals, the approximants are classified into four groups, while group 3 can be further divided into two subgroups. In figure 7b there is also a gap between $a$ = 14.0 to 15.0 Å, and four alloys including the present $Au_{51}Al_{35}Yb_{14}$ approximant fall into the gap.

Another aspect of size factor is matching between the radius of major and minor components, $r_S$ (S: small) and $r_L$ (L: large), respectively. This matching is due to geometrical condition for the construction of the triple shell structure with the dodecahedron, icosahedron and icosidodecahedron. From the geometrical consideration [3, 23] of close packing of two kinds of spheres the ideal ratio is

$$\frac{r_L}{r_S} = \left(\frac{10 + 5^{\frac{1}{2}} + 2 \times 3^{\frac{1}{2}}}{3}\right)^{\frac{1}{2}} - 1 \approx 1.288.$$

For the quasicrystal $Au_{51}Al_{34}Yb_{15}$, $r_S = r_{Au-Al}$ = 1.438 Å and then $r_L$ is expected to be 1.852 Å. However the trivalent Yb is too small, and contrary divalent Yb is too large. This consideration is valid not only for the quasicrystal but also for the approximant.

It is noted that Kashimoto [34] measured temperature dependence of magnetic susceptibility of the $Au_{51}Al_{34}Yb_{15}$ quasicrystal. Effective magnetic moment of Yb was measured to be 3.6$\mu_B$ that indicates the intermediate-valence state. Furthermore Watanuki *et al.* [35] has recently measured XANES (X-ray absorption near edge structure) directly indicating the occurrence of the intermediate-valence state at ambient pressure. These experimental results are consistent with the above argument on the atomic size.

In this quasicrystal Cd was replaced by smaller atoms Au-Al, and this substitution may cause the change of the valence of Yb to the direction of the intermediate-valence state. However, this situation seems opposite comparing with the case of Ga (or Al) substitution in $YbCu_5$ [9], where Ga (or Al) is larger than Cu. This apparent discrepancy suggests that the size effect on Yb valence is not simple and as the case may be. It is also necessary to study possible role of the chemical ordering of Au and Al for the appearance of the intermediate-valence state.

5. **Conclusion**

In this paper the formation of the icosahedral quasicrystal and its approximant was reported in Au-Al-Yb alloys. Both phases form at very similar compositions, and the quasicrystal is regarded as metastable. The relationship between the lattice parameters $a_{6D}$ and $a$ as well as the resemblance in alloy compositions suggest their close structural similarity. Comparing with other Tsai-type quasicrystal alloys, the present system is characterized by the combination of smaller major component Au-Al, and larger minor component Yb. This special combination may be one reason why the intermediate-valence state appears in the Au-Al-Yb quasicrystal.

**Acknowledgements**

The authors wish to express their thanks to T. Watanuki for interesting discussions. They also thank N. Miyazaki for his help and advice in the use of EPMA.

Figure captions

Figure 1    Powder X-ray diffraction pattern of as-cast $Au_{49}Al_{34}Yb_{17}$ specimen measured by Cu Kα radiation. Bars denote the expected positions of reflections for an icosahedral quasicrystal with 6-dimensional lattice parameter $a_{6D}$ = 7.448 Å. Corresponding six indices are inserted. An arrow at $2\theta$ = 30.02 ° indicates an unidentified peak.

Figure 2    Selected-area electron diffraction patterns of icosahedral quasicrystal formed in as-cast $Au_{49}Al_{34}Yb_{17}$ specimen. (a) 2-fold, (b) 3-fold and (c) 5-fold.

Figure 3    Powder X-ray diffraction pattern of $Au_{49}Al_{36}Yb_{15}$ specimen annealed at 700 ºC for 24 h. Calculated intensity, measured intensity, peak positions of the approximant, and deviation are presented in this order.

Figure 4    Selected-area electron diffraction patterns of approximant formed in annealed $Au_{49}Al_{36}Yb_{15}$ specimen. (a) [001], (b) [101] and (c) [111].    Reflections indicated by letters are A: 6 0 0, B: 0 6 0, C: 3 0 $\bar{3}$, D: 0 6 0, E: 3 0 $\bar{3}$, F: $\bar{3}$ 6 $\bar{3}$.

Figure 5    Structure model of Au-Al-Yb approximant in [1 0 0] projection.

Figure 6    (a) Framework of Yb atoms in [1 0 0] projection. *O*: octahedron, *P*: pyramid, *I*: icosahedron and *RP*: rhombus prism. A unit cell is drawn by thin lines. (b) ~ (e) atomic decoration of the octahedron, pyramid, icosahedron and rhombus prism, respectively.

Figure 7    Linear relationships between lattice parameters and average atomic radii.
 (a) Tsai-type quasicrystals and (b) Corresponding approximants. In (a), group 1 consists of $Cu_{46}Al_{38}Sc_{16}$ [15] and $Cu_{48}Ga_{34}Mg_{3}Sc_{15}$ [16,17]. Group 2: $Zn_{88}Sc_{12}$ [10], $Zn_{81}Mg_{4}Sc_{15}$ [11], $Zn_{72}Cu_{12}Sc_{16}$ [12], $Zn_{74}Mn_{10}Sc_{16}$, $Zn_{77}Fe_{7}Sc_{16}$, $Zn_{78}Co_{6}Sc_{16}$, $Zn_{74}Ni_{10}Sc_{16}$ [14, 23], $Zn_{75}Pd_{9}Sc_{16}$, $Zn_{74}Ag_{10}Sc_{16}$, $Zn_{74}Pt_{10}Sc_{16}$, $Zn_{74}Au_{10}Sc_{16}$ [13, 23]. Group 3: $Au_{44.2}In_{41.7}Ca_{14.1}$ [21], $Cd_{65}Mg_{20}Y_{15}$, $Cd_{65}Mg_{20}Gd_{15}$, $Cd_{65}Mg_{20}Tb_{15}$, $Cd_{65}Mg_{20}Dy_{15}$, $Cd_{65}Mg_{20}Ho_{15}$, $Cd_{65}Mg_{20}Er_{15}$, $Cd_{65}Mg_{20}Tm_{15}$, $Cd_{65}Mg_{20}Lu_{15}$ [18], $Ag_{42}In_{42}Ca_{16}$ and $Ag_{42}In_{42}Yb_{16}$ [5]. Group 4: $Cd_{85}Ca_{15}$, $Cd_{84}Yb_{16}$ [1, 2], $Cd_{65}Mg_{20}Ca_{15}$ and $Cd_{65}Mg_{20}Yb_{15}$ [4].    In (b),    group 1 consists of $Cu_{45.0}Al_{40.5}Sc_{14.5}$ and $Cu_{3.7}Ga_{2.3}Sc$ [31].    Group 2: $Zn_{6}Sc$ and $Zn_{16.5}Cu_{1.48}Sc_{3}$ [12]. Group 3a: $Ag_{46.4}In_{39.7}Gd_{13.9}$, $Ag_{42.2}In_{42.6}Tm_{15.2}$, $Au_{49.7}In_{35.4}Ce_{14.9}$, $Au_{47.2}In_{37.2}Gd_{15.6}$ [32], $Au_{56.7}In_{29.3}Ca_{14.0}$ [21], Group 3b: $Ag_{48}In_{38}Ce_{14}$, $Ag_{47}In_{39}Pr_{14}$, $Ag_{40}In_{46}Yb_{14}$ [32], $Cd_{6}Y$, $Cd_{6}Nd$, $Cd_{6}Sm$, $Cd_{6}Gd$, $Cd_{6}Dy$ [29].    Group 4: $Cd_{6}Ca$ and $Cd_{6}Yb$ [29].

Table 1. Structure model of Au-Al-Yb 1/1 cubic approximant. Occupancies of the site M7 and M8 are 16.7 %. and 92 %, respectively.   *:   24g site. See reference [21] for details.

| Site | Atom | Set | x | y | z | B (Å²) | Au-Ga-Ca |
|---|---|---|---|---|---|---|---|
| M1 | 0.728Au + 0.272Al | 48h | 0.3428(1) | 0.1966(2) | 0.1077(1) | 1.92(5) | Au/Ga2 (0.75/0.25) |
| M2 | 0.380Au + 0.620Al | 24g | 0 | 0.2503(4) | 0.0856(3) | 3.1(2) | Au5 (0.25) + Ga5 (0.75) |
| M3 | 0.882Au + 0.118Al | 24g | 0 | 0.5954(1) | 0.6488(2) | 1.72(6) | Au1 |
| M4 | 0.956Au + 0.044Al | 16f | 0.1525(1) | --- | --- | 2.21(8) | Au4 |
| M5 | 0.033Au + 0.967Al | 12e | 0.192(1) | 0 | 1/2 | 1.3(3) | Ga3 |
| M6 | 0.411Au + 0.589Al | 12d | 0.3980(6) | 0 | 0 | 2.4(2) | Au/Ga6 (0.40/0.60) |
| M7 | 0.078Au + 0.089Al | 48h | 0.045(2) | 0.027(2) | 0.095(1) | 6.3(9) | Au7 (0.077) + Ga7 (0.025)* |
| M8 | 0.92Al | 8c | 1/4 | 1/4 | 1/4 | 1.2(7) | Ga8 |
| Yb | Yb | 24g | 0 | 0.1894(2) | 0.3035(2) | 0.57(5) | Ca |

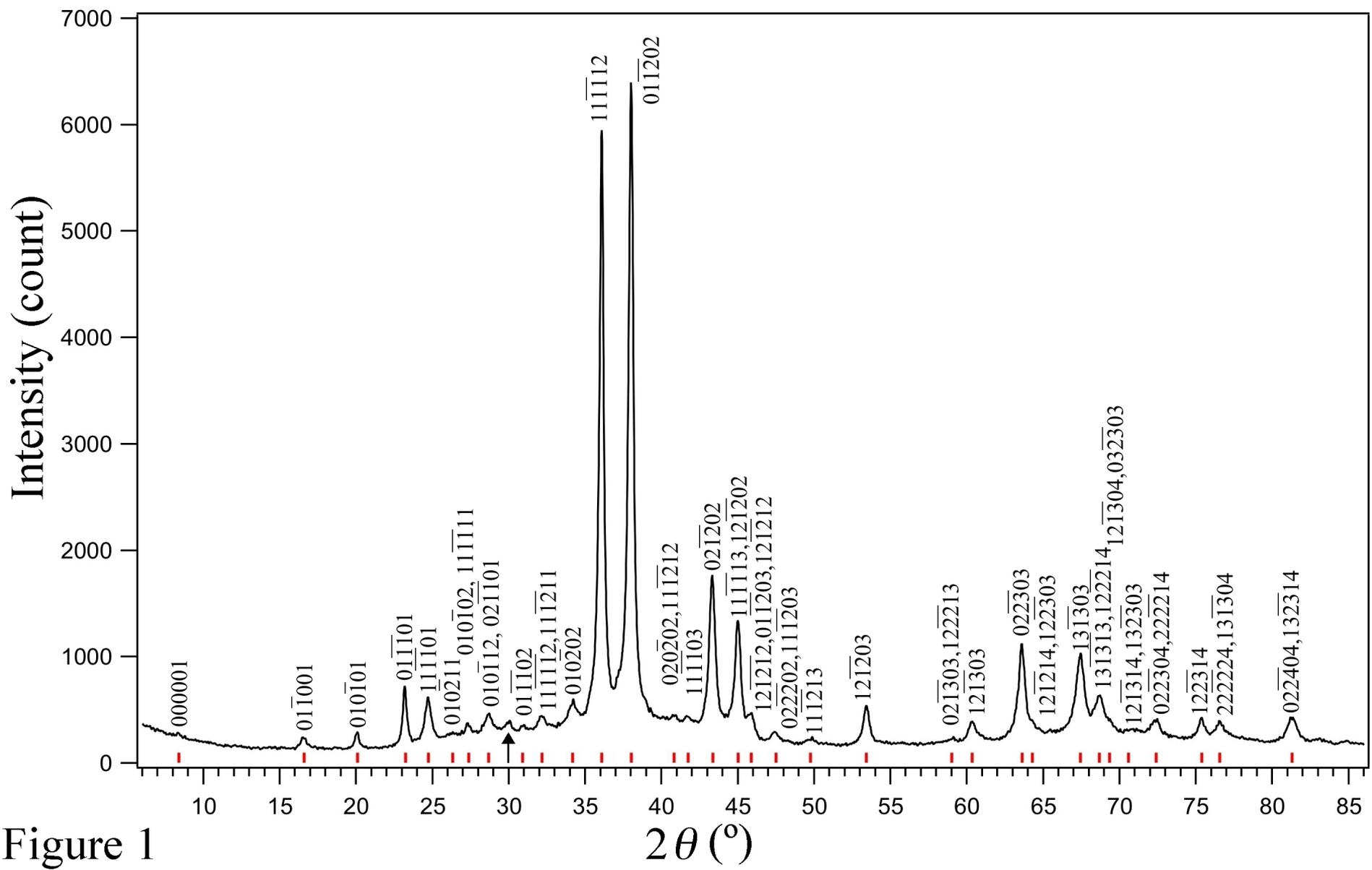

Figure 1

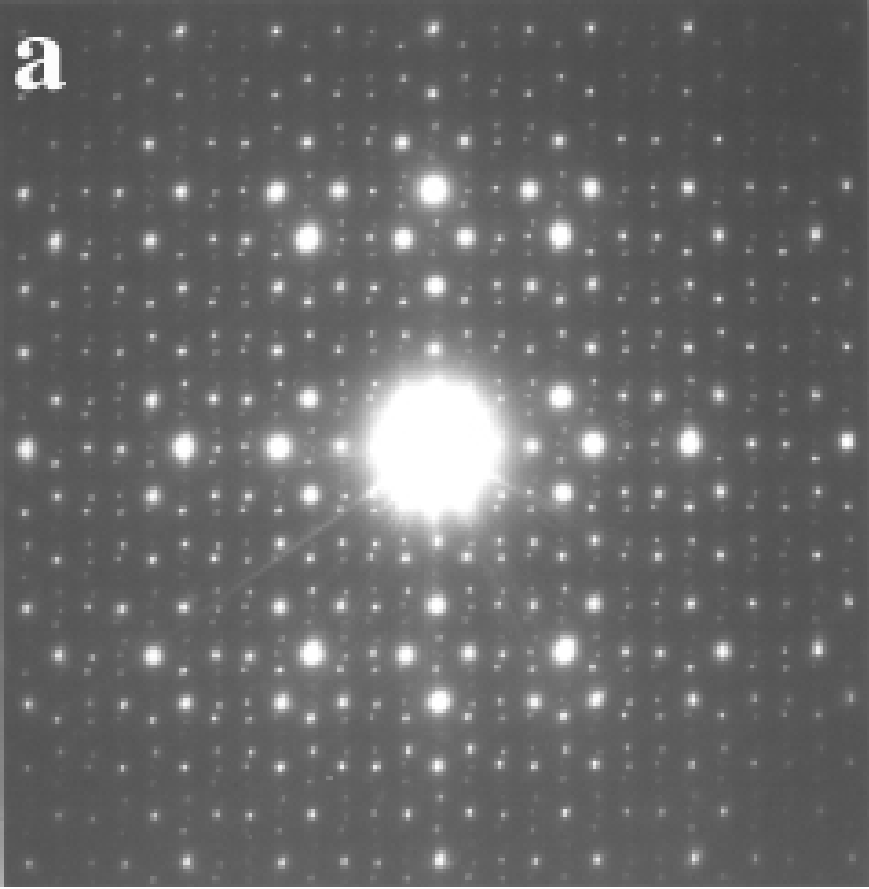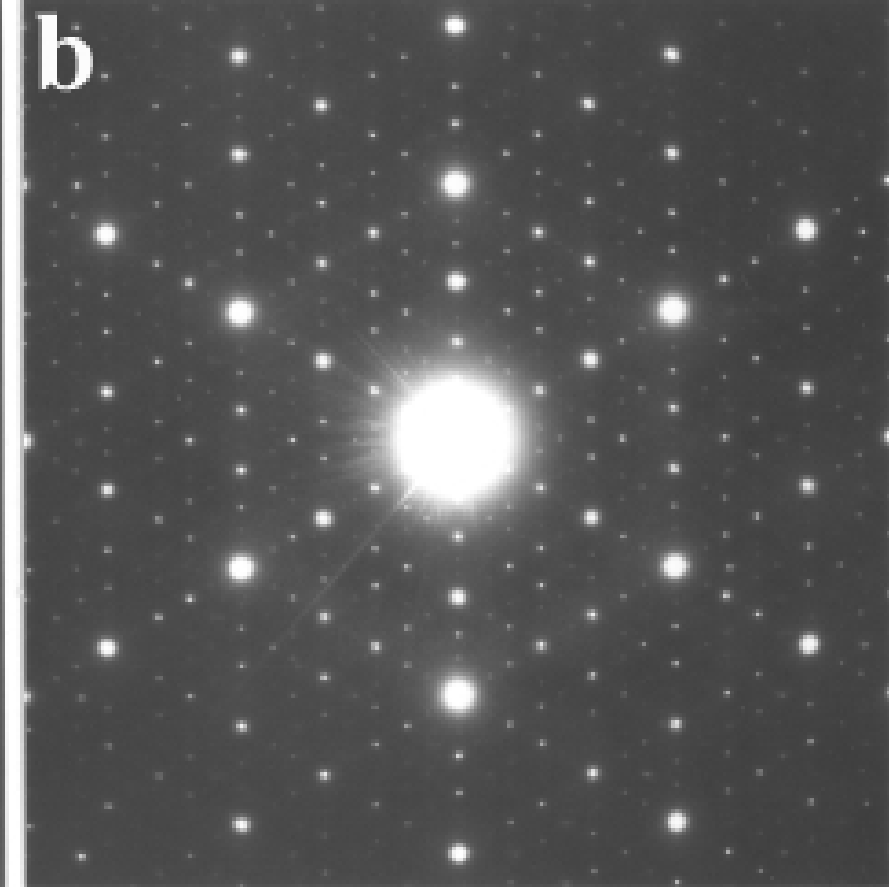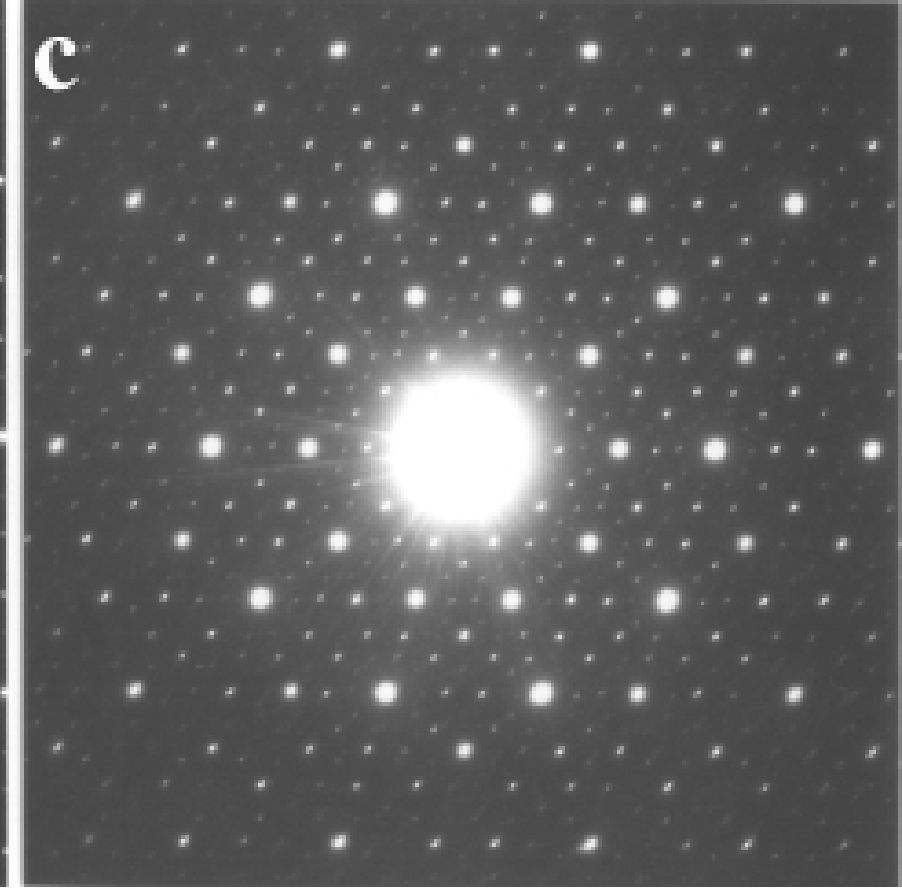

Figure 2

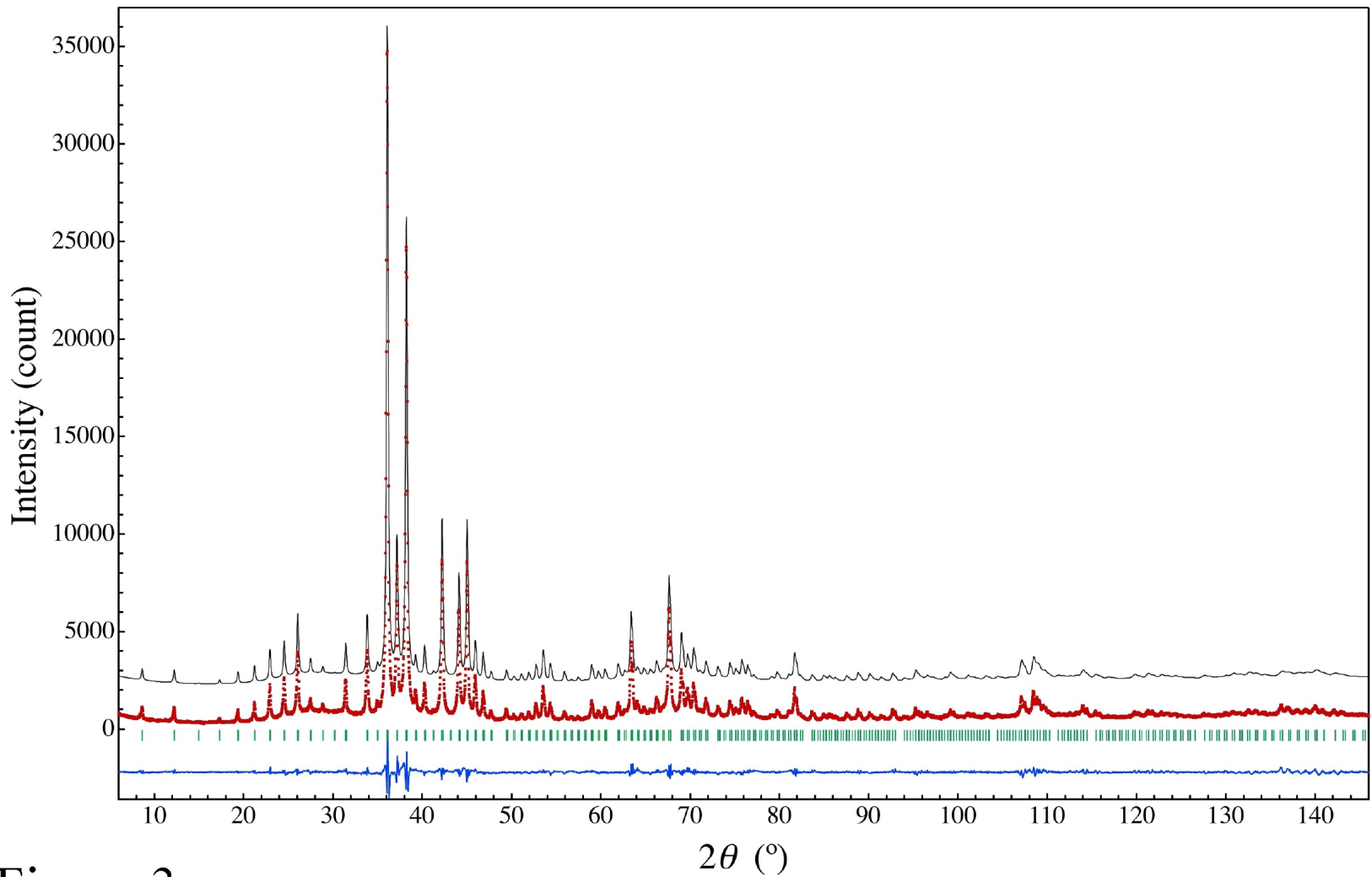

Figure 3

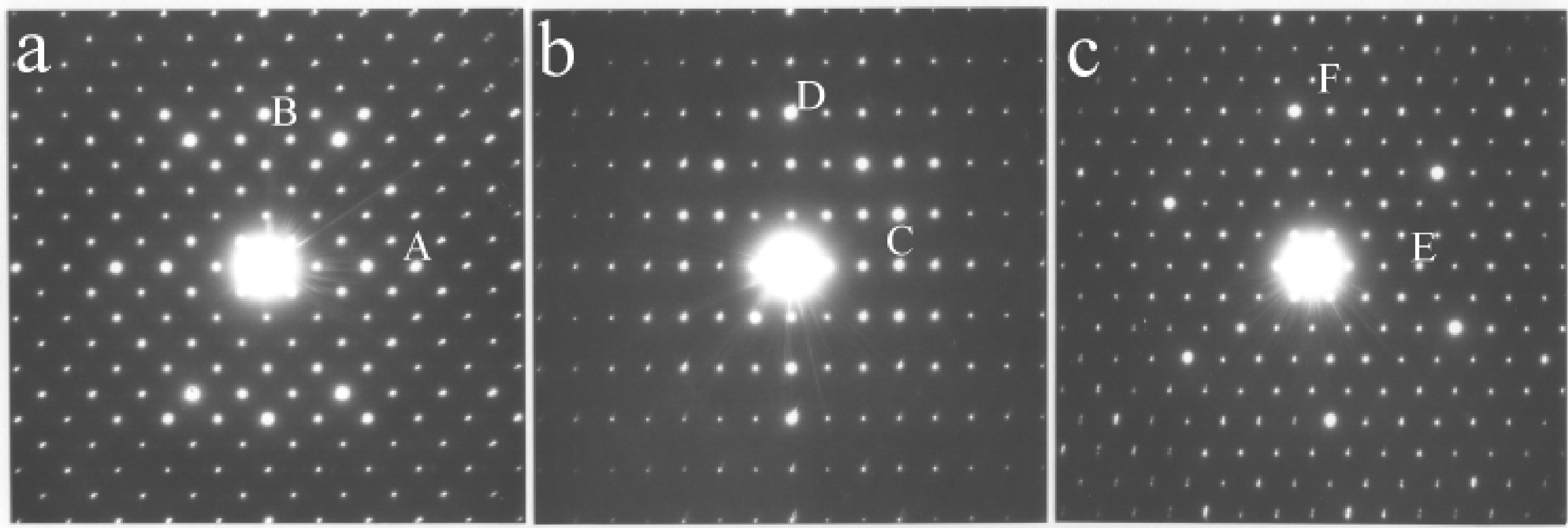

Figure 4

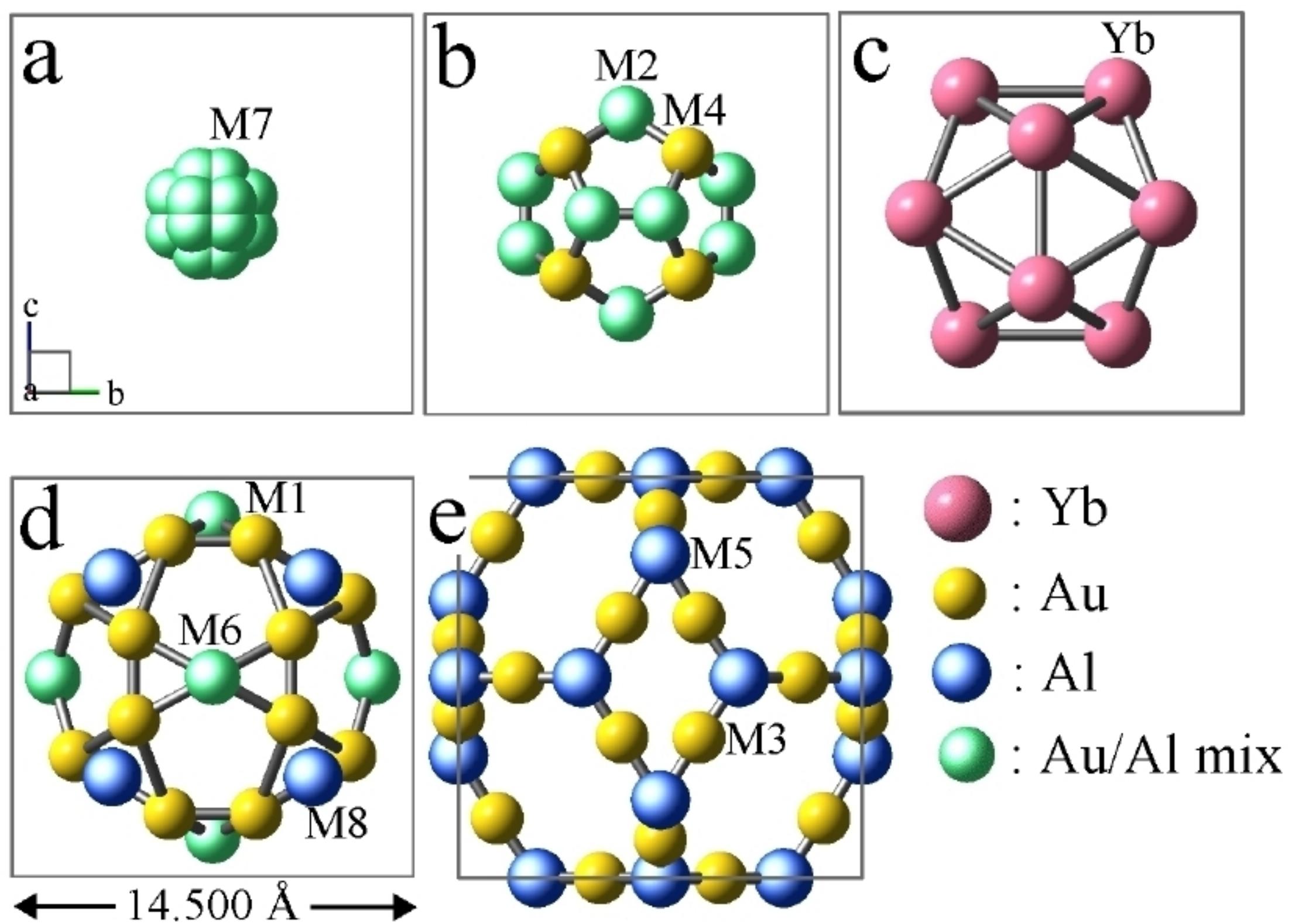

Figure 5

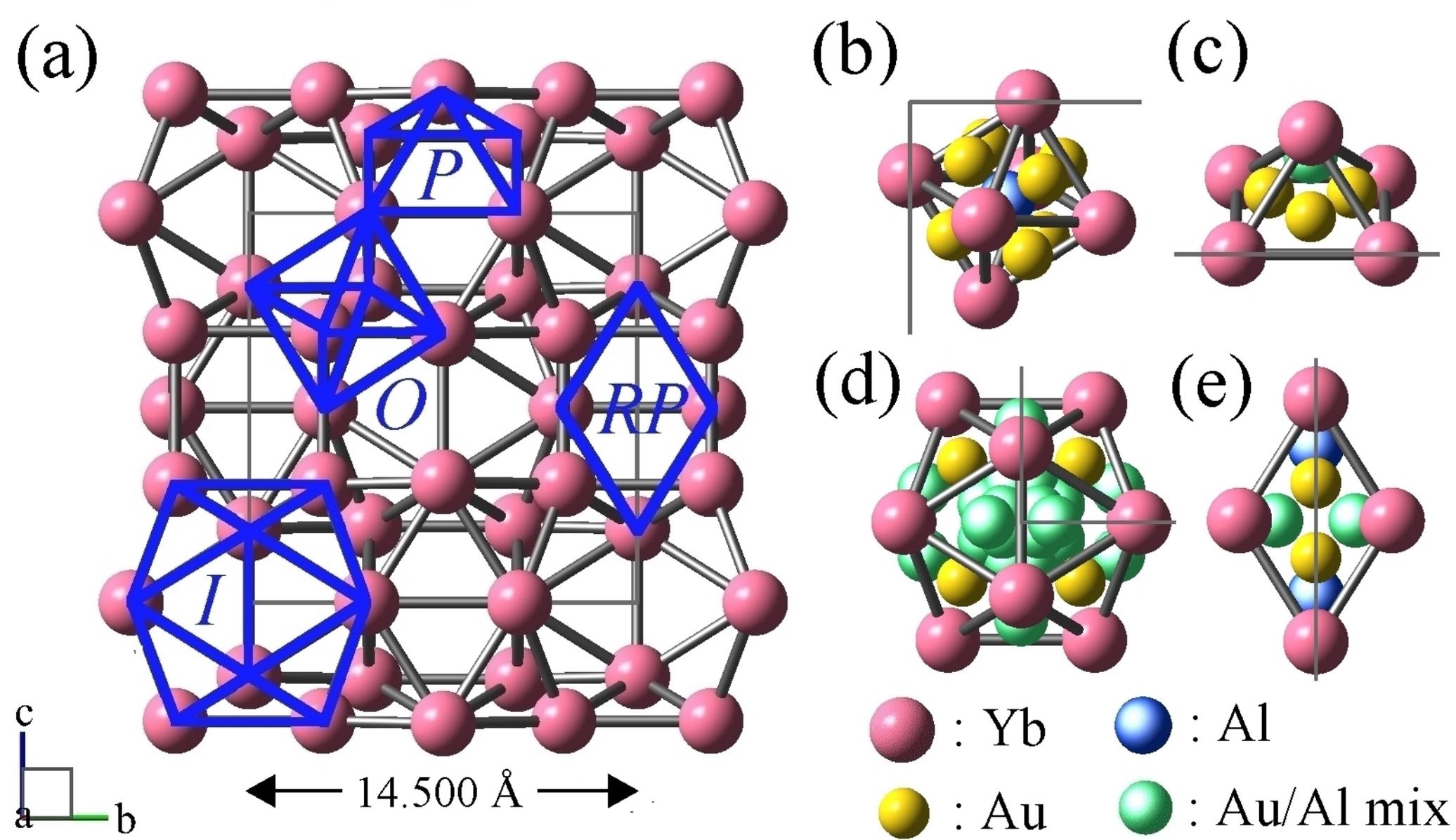

Figure 6

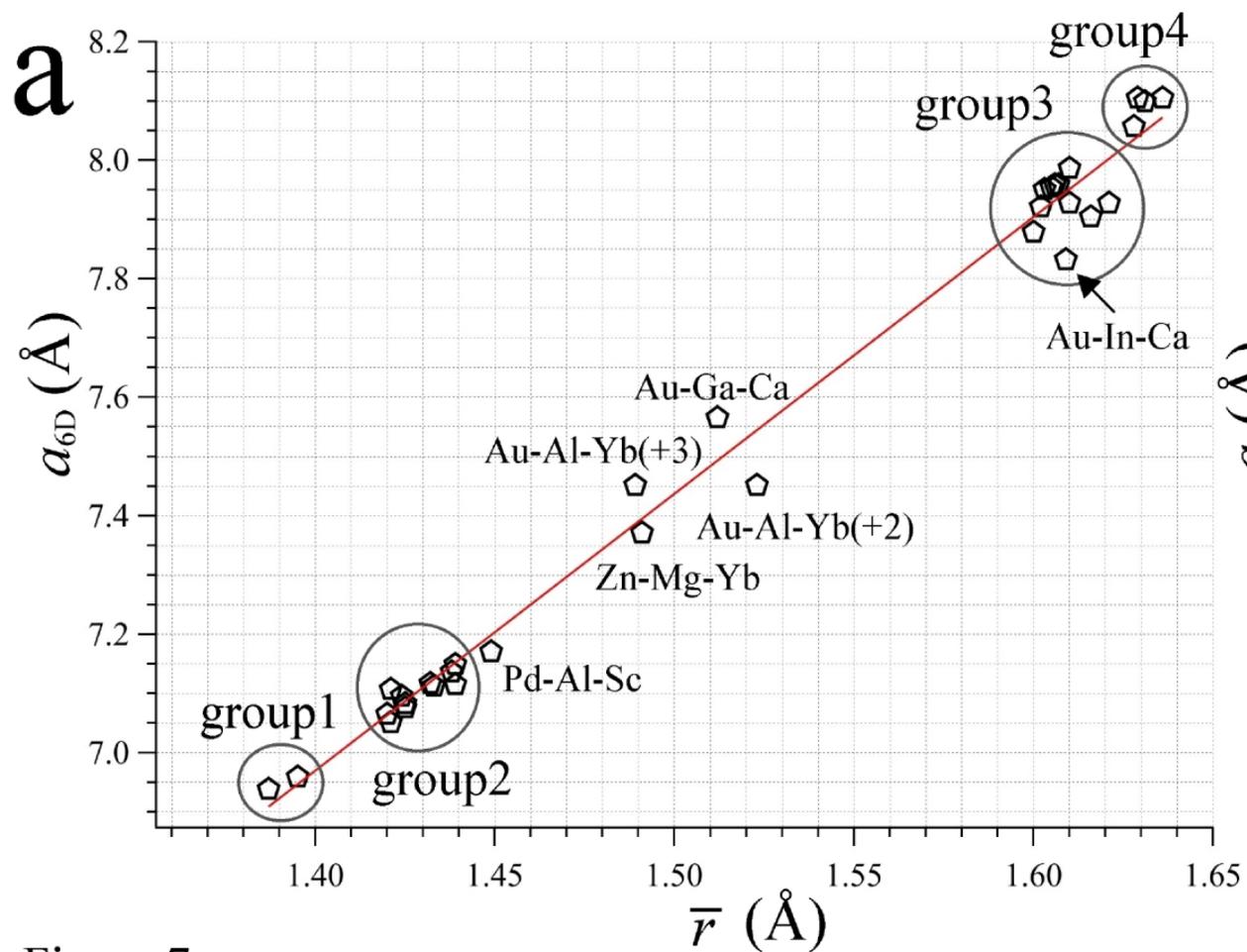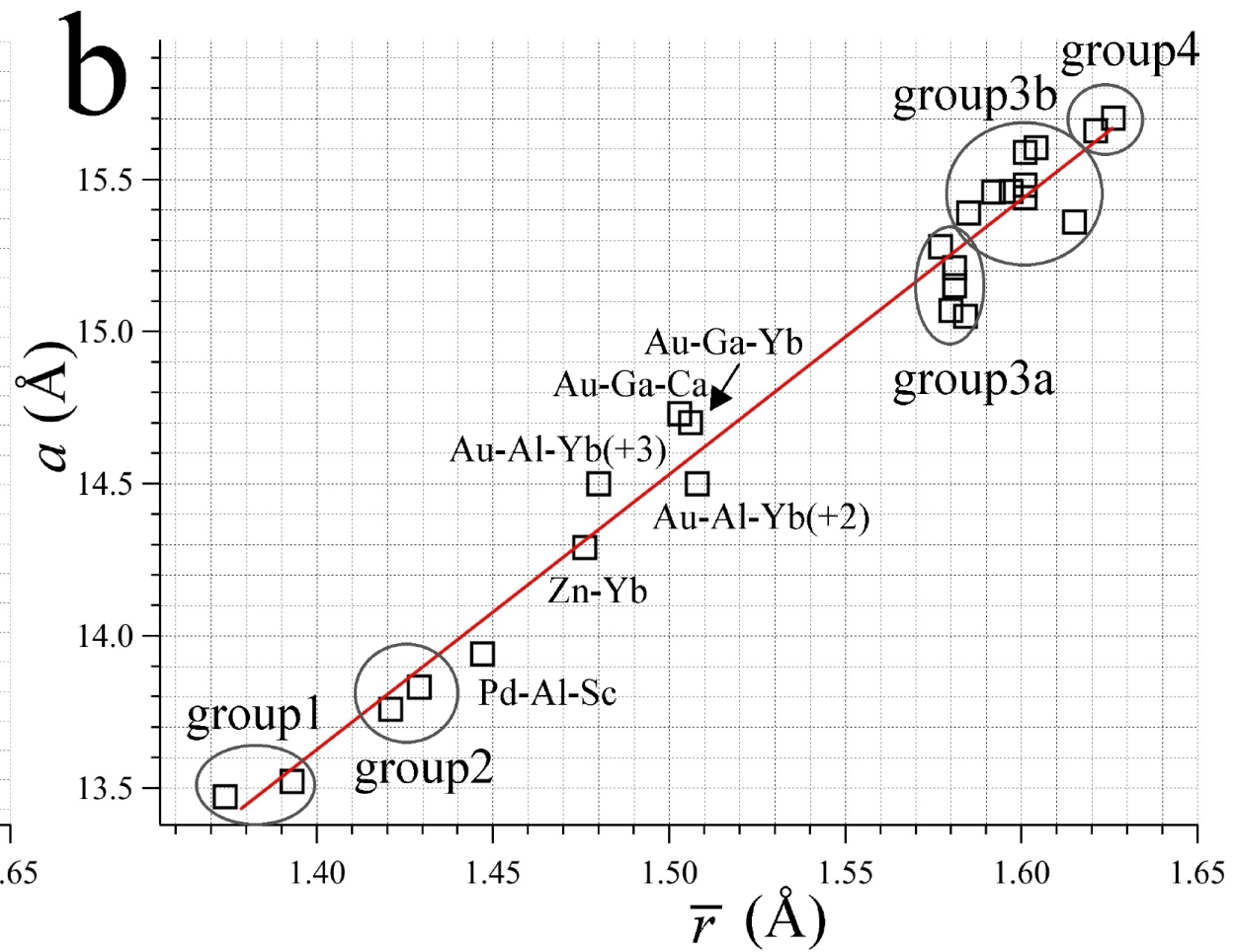

Figure 7